\documentclass[epj]{webofc}
\usepackage[utf8]{inputenc}
\usepackage[varg]{txfonts}   
\usepackage{booktabs}
\usepackage{xcolor}
\definecolor{darkred}{rgb}{0.4,0.0,0.0}
\definecolor{darkgreen}{rgb}{0.0,0.4,0.0}
\definecolor{darkblue}{rgb}{0.0,0.0,0.4}
\usepackage[bookmarks,linktocpage,colorlinks,
    linkcolor = darkred,
    urlcolor  = darkblue,
    citecolor = darkgreen]{hyperref}
\usepackage{subfigure}
\wocname{EPJ Web of Conferences}
\woctitle{Lattice2017}
\begin{document}
%
\selectlanguage{english}
\title{%
Atiyah-Patodi-Singer index theorem for domain-wall fermion Dirac operator
}
\author{%
\firstname{Hidenori} \lastname{Fukaya}\inst{1}\fnsep\thanks{Acknowledges financial support  in part by the Japanese Grant-in-Aid for
Scientific Research(No. JP26247043).} \and
\firstname{Tetsuya} \lastname{Onogi}\inst{1}\fnsep\thanks{Speaker, 
\email{onogi@phys.sci.osaka-u.ac.jp}} \and
\firstname{Satoshi}  \lastname{Yamaguchi}\inst{1}\fnsep\thanks{Acknowledges financial support  in part by the Japanese Grant-in-Aid for
Scientific Research(No. JP15K05054).}
}
\institute{%
Department of Physics, Osaka University, 1-1 Machikaneyama-cho, 
Toyonaka, 560-0043, Osaka, JAPAN
}
\abstract{%
Recently, the Atiyah-Patodi-Singer(APS) index theorem attracts attention for understanding physics on the surface of materials in topological phases. Although it is widely applied to physics, the mathematical set-up in the original APS index theorem is too abstract and general (allowing non-trivial metric and so on) and also the connection between the APS boundary condition and the physical boundary condition on the surface of topological material is unclear. For this reason, in contrast to the Atiyah-Singer index theorem, derivation of the APS index theorem in physics language is still missing. In this talk, we attempt to reformulate the APS index in a "physicist-friendly" way, similar to the Fujikawa method on closed manifolds, for our familiar domain-wall fermion Dirac operator in a flat Euclidean space. We find that the APS index is naturally embedded in the determinant of domain-wall fermions, representing the so-called anomaly descent equations.}
\maketitle
\section{Introduction}\label{intro}

Massive fermions in D-dimensional bulk spacetime with (D-1)-dimensional boundary attracts attention in various fields of physics. The domain-wall fermion in five-dimensional bulk spacetime is now widely used in lattice QCD to realize chiral fermions on the lattice \cite{Kaplan:1992bt, Shamir:1993zy} .  Another example is the topological insulators in three- or four-dimensional bulk spacetime which are nowadays the main focus of research in condensed matter physics.  

The common feature in these systems is the existence of a massless fermion excitation at the boundary in the case when the massive fermion at the bulk is in the symmetry protected topological (SPT) phase. One can show that massless fermion at the boundary is realized as a solution of the massive Dirac equation which is localized near the boundary. The emergence of such an edge-mode is not an accident but robust for symmetry reasons. The key is the cancellation of anomaly between the bulk contribution and the boundary contribution. For D=odd, 
it is the cancellation of perturbative gauge anomaly.  For D=even, it is the cancellation of time reversal symmetry anomaly (T-anomaly).  While the former case is known as the Callan-Harvey mechanism \cite{Callan:1984sa}, the latter is less familiar in particle physics but is known to be related to the mathematical theorem called Atiyah-Patodi-Singer (APS) index theorem \cite{Atiyah:1975jf, Atiyah:1976jg, Atiyah:1980jh}
, which is an extension of Atiyah-Singer index theorem \cite{Atiyah:1963zz, Atiyah:1968mp} for massless Dirac operator on  manifolds with boundaries. In this work, we study the D=4 case and try to give a more physically intuitive derivation of the APS index  in the context of T-anomaly cancellation in topological insulators.

Before explaining our goal, let us introduce the discussion by Witten on the relation between the T-anomaly cancellation and APS index theorem \cite{Witten:2015aba}. 
Consider a massive fermion in 4-dimensional spacetime manifold $X$ without boundary first.  What is the SPT phase in this case? 
The simplest characterization is made by the effective action $S_{\rm eff}[A]$,  after integrating out the massive fermion,  as a functional of the 
external gauge field $A_\mu(x)$ that is introduced as a probe:
\begin{eqnarray}
Z = \exp(-S_{\rm eff}[A]) = \mbox{Det}\left(\frac{D^{4d}+M}{D^{4d}+\Lambda}\right), 
\end{eqnarray}
where $D^{4d}$ is the gauge covariant 4-dimensional Dirac operator, $M$ is the mass of the fermion and $\Lambda$ is the 
mass of the Pauli-Villars regulator field.  In general, the imaginary part of the effective action or the phase of the partition function is given 
as 
\begin{eqnarray}
Z = |Z| \exp(i \theta P),  
\end{eqnarray} 
where $P$ is an integer defined by the instanton number as
\begin{eqnarray}
P= \frac{1}{32\pi^2} \int_X d^4x \epsilon^{\mu\nu\alpha\beta}\mbox{tr}[F_{\mu\nu}(x)F_{\alpha\beta}(x)].
\end{eqnarray} 
Since the Pauli-Villars regulator mass does not break the time reversal (T) symmetry in four dimensions, the effective action also 
does not break the T-symmetry, one finds that only $\theta=0$ or $\theta=\pi$ is allowed.  And SPT phase corresponds to the case 
with $\theta=\pi$, which happens when the physical mass $M$ has an opposite sign with the regulator mass $\Lambda$.

When the insulator in SPT phase has a boundary, i.e. it lives on the manifold $X$ with the boundary $Y$,  
the effective action is modified due to the boundary effect. Although it is difficult to exactly separate the effective action into "the bulk contribution" and "the boundary contribution", one naively expects that the bulk part should give 
\begin{eqnarray}
Z_{\rm bulk} \approx |Z_{\rm bulk}| \exp(i \theta P),  
 & & P= \frac{1}{32\pi^2} \int_X d^4x \epsilon^{\mu\nu\alpha\beta}\mbox{tr}[F_{\mu\nu}(x)F_{\alpha\beta}(x)].
\end{eqnarray} 
The boundary contribution is expected to be described the massless edge-modes on the boundary which is described by the 
effective three dimensional theory on the boundary $Y$  so that 
\begin{eqnarray}
Z_{\rm boundary} \approx \mbox{Det}(D^{3d}) = \prod_i \frac{\lambda_i}{\lambda_i + i \Lambda}
= \left|Z_{\rm boundary}\right| \exp\left( -i\frac{\pi}{2}\eta(i D^{3d})\right), 
\end{eqnarray} 
where $\lambda_i$ is the eigenvalue of the hermitian operator $i D^{3d}$ and $\eta(i D^{3d})$ 
is the eta-invariant in three dimension defined by
\begin{eqnarray}
\eta(i D^{3d})= \lim_{s\rightarrow 0} \sum_k \mbox{sign}(\lambda_k) |\lambda_k|^{-s}.
\end{eqnarray}
Combining "bulk contribution" and "boundary contribution" one naively obtains
\begin{eqnarray}
Z = Z_{\rm bulk} Z_{\rm boundary}  \approx |Z| \exp\left(  i \pi \left( P - \frac{\eta(D^{3d})}{2} \right)\right)
\end{eqnarray}
Since we know that the Hamiltonian of the insulator does not break the T symmetry even with the boundary, 
the T-anomaly should be canceled. Thus the question is whether the quantity $\mathcal{I}$ defined by 
\begin{eqnarray}
\mathcal{I} \equiv P - \frac{\eta(D^{3d})}{2}  
\end{eqnarray}
becomes integer or not, which is actually answered by the APS index theorem as explained below.


Let us consider the  four-dimensional Dirac operator of the following form with U(1) or SU(N) gauge 
field in the $A_4=0$ gauge
\begin{eqnarray}
  D^{4d} &=& \gamma_4(\partial_4 + A),
\end{eqnarray}
where $A=\gamma_4 \sum_{i=1}^3 \gamma_i D_i$ with covariant derivative $D_i=\partial_i+iA_i$,
is a Hermitian operator.  We consider a  four-dimensional flat manifold $X$ extending in the region $x_4>0$,
with a three-dimensional boundary $Y$ at $x_4=0$.

The boundary condition of the fermion fields is taken as 
\begin{eqnarray}
\left.\frac{A+|A|}{2}\phi\right|_{x_4=0} = 0, 
\end{eqnarray}
which is called as the APS boundary condition.
This is a non-local boundary condition which keeps the anti-Hermiticity 
of  $D$. This boundary condition also preserve the chirality. 

Atiyah-Patodi-Singer \cite{Atiyah:1975jf} have proven that the index of $D^{4d}$
which is defined by the difference of the numbers of eigenmodes with positive and negative chirality 
can be given as
\begin{eqnarray}
\mbox{Index of } D^{4d} = n_+ - n_- = \mathcal{I} \equiv  P -\frac{\eta(i D^{3d})}{2}.
\end{eqnarray}

Therefore, APS index theorem tells us that $\mathcal{I}$ must be an integer, which 
suggests that the T-anomaly in the partition function defined by the product of "bulk contribution" and 
"boundary contribution" is canceled. This is an integral expression of (a part of) the anomaly descent equations.

Although looks reasonable, the above argument is unsatisfactory on several points which give rise to the following questions:
1) It is not an exact calculation based on the full microscopic theory but a discussion based on the low energy effective theory 
which deals with the main contributions to the T-anomaly. This is somewhat similar in spirit with the anomaly 
matching condition by 't Hooft.  Is there any rigorous proof of T-anomaly cancellation?
2) The APS index is an index of massless Dirac operator with non-local boundary condition. However, topological 
insulator considers massive fermion with local boundary condition. Although useful, the APS index theorem seems 
to have no physical relation with the domain-wall fermion or topological insulator with boundary. 
Is there any different formulation of the index, which is directly related to domain-wall fermion?
3) In the AS index theorem, Fujikawa method gave a simple and physically intuitive derivation.
Is there any simple derivation of APS index theorem using Fujikawa method \cite{Fujikawa:1979ay}?

In order to answer these questions, we reformulate the index theorem based on the 
domain-wall fermion Dirac operator having four-dimensional bulk space-time 
and give a physically-intuitive proof of the index theorem using Fujikawa method ~\cite{Fukaya:2017tsq}. 
\footnote{For the special case with two boundaries at $t=-\infty$ and $t=+\infty$ where $t$ dependence of the 
gauge field is negligible near the boundaries, APS index theorem is derived by path integral method \cite{AlvarezGaume:1984nf}.}

\section{Index for domain wall Dirac operator}\label{sec:upload}

The domain-wall fermion Dirac operator is defined by
\begin{eqnarray}
  D_{DW} = D + M\epsilon(x_4),\;\;\epsilon(x_4) = {\rm sign}\; x_4.
\end{eqnarray}
where the mass term flips its sign across the domain-wall located at $x_4=0$.
Here and in the following, we take $M$ to be positive.
In lattice gauge theory, we often consider the domain-wall fermion determinant
together with a Pauli-Villars field,
\begin{eqnarray}
  \label{eq:DWdet}
  \det \frac{D + M\epsilon(x_4)}{D-M},
\end{eqnarray}
to cancel the bulk mode effects in the region $x_4<0$,
while the $x_4>0$ region gives a non-trivial contribution.
Note here that fermion field is defined in the
whole $-\infty<x_4<\infty$ region and
no boundary condition is imposed on it.
This determinant provides a good model to describe
fermions in a topological insulator located in the $x_4>0$ region,
surrounded by a normal insulator sitting in the $x_4<0$ region.
As we will explicitly show, the edge-localized modes appear at the boundary $x_4=0$,
and play a crucial role in the definition of the index.

The determinant Eq.~(\ref{eq:DWdet}) 
is real,  due to the ``$\gamma_5$ Hermiticity'':
\begin{eqnarray}
\label{eq:DWdet2}
  \det \left[(D + M\epsilon(x_4))(D - M)^{-1}\right]
  &=& \det \left[(D^\dagger + M\epsilon(x_4))(D^\dagger - M)^{-1}\right] \nonumber\\
  &=& \left|\det \left[(D^\dagger + M\epsilon(x_4))(D^\dagger - M)^{-1}\right]\right| (-1)^{\mathcal{I}},
\end{eqnarray}
where $\mathcal{I}$ is an integer determining the sign of the determinant.
In fact, we will explicitly show that this integer $\mathcal{I}$
is equivalent to the APS index.
A similar statement is found in \cite{Witten:2015aba},
but no explicit bulk fermion determinant
nor its boundary condition is given.
The ``outside'' of our target domain is not mentioned, either.
As will be shown below, we need no massless Dirac operator nor
non-local APS boundary condition for the ``new'' index.

Our new index $\mathcal{I}$ is formally defined by
a regularized eta-invariant of the Hermitian operator
$H_{DW} = \gamma_5 (D + M\epsilon(x_4))$:
\begin{eqnarray}
\mathcal{I} &\equiv& \frac{\eta(H^{reg}_{DW})}{2}
= \frac{1}{2}\eta(H_{DW})- \frac{1}{2}\eta(H_{PV}),
\end{eqnarray}
where we employ the Pauli-Villars regularization
with another Hermitian operator $H_{PV} = \gamma_5 (D-M)$.
This definition coincides with the exponent appeared in Eq.(\ref{eq:DWdet2}) as 
\begin{eqnarray}
  \det \frac{D + M\epsilon(x_4)}{D-M} =
  \det \frac{iH_{DW}}{iH_{PV}}
  =
  \prod_{\lambda_{DW}} i\lambda_{DW} / \prod_{\lambda_{PV}} i\lambda_{PV}
&\propto&  \exp\left(\frac{i\pi}{2}\left(\sum_{\lambda_{DW}}{\rm sign} \lambda_{DW}
  -\sum_{\lambda_{PV}}{\rm sign} \lambda_{PV}\right)\right) \
  \nonumber\\
  &=& (-1)^{\frac{1}{2}\eta(H_{DW})- \frac{1}{2}\eta(H_{PV})}.
\end{eqnarray}
In the following, we compute the two eta-invariants $\eta(H_{DW})$ and $\eta(H_{PV})$ separately, by
introducing another regularization using the (generalized) $\zeta$ function
(we simply call the $\zeta$-function regularization).
This double regularization is not theoretically needed but
simplifies the computation and clarifies the role of
the Pauli-Villars fields.
In fact, we will see that $\eta(H_{DW})/2$ alone gives
only a ``half'' of the (bulk contribution of) total APS index,
to which another ``half'' is provided by $\eta(H_{PV})/2$.

\subsection{$\eta(H_{PV})$ }

The Pauli-Villars part $\eta(H_{PV})$ is easily obtained by the standard Fujikawa method,
which coincides with the AS index,
\begin{eqnarray}
  \label{eq:massiveFujikawaclosed}
  \eta(H_{PV}) &=& \lim_{s\to 0}{\rm Tr}\frac{H_{PV}}{(\sqrt{H_{PV}^2})^{1+s}}
  = \lim_{s\to 0}\frac{1}{\Gamma\left(\frac{1+s}{2}\right)}
  \int_0^\infty dt t^{\frac{s-1}{2}}{\rm Tr}H_{PV} e^{-t H_{PV}^2}\nonumber\\
  &=& -\frac{1}{\sqrt{\pi}}\int_0^\infty dt' t'^{-\frac{1}{2}}{\rm Tr} \gamma_5\left(1-\frac{D}{M}\right) e^{-t' D^\dagger D/M^2 }e^{-t'},
  \nonumber\\&=& -\frac{1}{32\pi^2} \int d^4x\; \epsilon_{\mu\nu\rho\sigma}{\rm tr}_cF^{\mu\nu}F^{\rho\sigma} +\mathcal{O}(1/M^2).
\end{eqnarray}
This simple computation tells us that the masslessness is
not required to define the AS index.
As will be shown below, this is also true in the case with boundary.

\subsection{$\eta(H_{DW})$ }

Now our goal is to compute
the eta-invariant,
\begin{eqnarray}
  \eta(H_{DW}) = \lim_{s\to 0}\left[{\rm Tr}(M\gamma_5\epsilon(x_4))\left(\sqrt{H_{DW}^2}\right)^{-1-s}
  +{\rm Tr}(\gamma_5 D)\left(\sqrt{H_{DW}^2}\right)^{-1-s}\right].  
\end{eqnarray}
As shown below, the second term includes contribution from
massless edge-localized modes, and it is non-local in general.
Following the general strategy to compute the ``local'' part
of the phase of the odd-dimensional massless fermion determinant \cite{AlvarezGaume:1984nf},
we consider a one-parameter family of gauge fields $uA_\mu$,
and take a $u$-derivative and integrate it again,
\begin{eqnarray}
  \int_0^1 du \frac{d}{du}\left[{\rm Tr}(H_{DW}(u)-M\gamma_5\epsilon(x_4))\left(\sqrt{H_{DW}(u)^2}\right)^{-1-s}\right]
    \nonumber\\
    =\int_0^1 du {\rm Tr}\left[-s\frac{d}{du}H_{DW}(u)\left(\sqrt{H_{DW}(u)^2}\right)^{-1-s}
    -\frac{d}{du}\left(\gamma_5 M\epsilon(x_4)\left(\sqrt{H_{DW}(u)^2}\right)^{-1-s}\right)\right],
\end{eqnarray}
where $H_{DW}(u)$ is the corresponding domain-wall fermion Dirac operator at $u$.
This procedure allows us to compute the eta-invariant
up to an integer, which may depend on a winding number of gauge transformation
on the surface.
Using the formula
\begin{eqnarray}
  \frac{1}{(\sqrt{O^2})^{1+s}} =
   \frac{1}{\Gamma\left(\frac{1+s}{2}\right)}\int_0^\infty dt\; t^{\frac{s-1}{2}}e^{-t O^2},
\end{eqnarray}
for a Hermitian operator $O$, we will compute
\begin{eqnarray}
  \label{eq:goalDW}
  \eta(H_{DW}) &=& \lim_{s\to 0}\frac{1}{\Gamma\left(\frac{1+s}{2}\right)}\int_0^\infty dt\; t^{\frac{s-1}{2}}
  \lim_{M\to \infty}{\rm Tr}\left[\gamma_5\epsilon(x_4) e^{-t \frac{H_{DW}^2}{M^2}}\right]
  \nonumber\\&&
  +\int_0^1 du \lim_{s\to 0}\frac{1}{\Gamma\left(\frac{1+s}{2}\right)}\int_0^\infty dt\; t^{\frac{s-1}{2}}
  \lim_{M\to \infty}{\rm Tr}\left[-s\frac{dH_{DW}(u)}{du}\frac{e^{-t \frac{H_{DW}(u)^2}{M^2}}}{M}\right]
  \nonumber\\&&
  -\int_0^1 du \frac{d}{du}\left\{\lim_{s\to 0}\frac{1}{\Gamma\left(\frac{1+s}{2}\right)}\int_0^\infty dt\; t^{\frac{s-1}{2}}
  \lim_{M\to \infty} {\rm Tr}\left[\gamma_5\epsilon(x_4) e^{-t \frac{H_{DW}(u)^2}{M^2}}\right]\right\}.
\end{eqnarray}

In our computation, the Dirac representation for
the gamma matrices,
\begin{eqnarray}
  \gamma_{i=1,2,3}&=&\left(\begin{array}{cc}
     & \sigma_i\\
   \sigma_i & 
  \end{array}\right)=\tau_1\otimes \sigma_i,\;\;\;
  \gamma_4=\left(\begin{array}{cc}
    1_{2\times 2} &\\
   & -1_{2\times 2}
  \end{array}\right)=\tau_3\otimes 1_{2\times 2},
  \nonumber\\
  \gamma_5&=&-\gamma_1\gamma_2\gamma_3\gamma_4=\left(\begin{array}{cc}
    & i 1_{2\times 2} \\
   -i 1_{2\times 2} & 
  \end{array}\right)=-\tau_2\otimes 1_{2\times 2},  
\end{eqnarray}
is useful.
Our Hermitian Dirac operator is then expressed by
\begin{eqnarray}
  H_{DW}
  &=& \left(\begin{array}{cc}
     & -i(\partial_4-M\epsilon(x_4)) \\ 
    -i(\partial_4+M\epsilon(x_4))&   
  \end{array}\right)
  +\left(\begin{array}{cc}
    -iD^{\rm 3D} &\\
   & iD^{\rm 3D}
  \end{array}\right).
\end{eqnarray}


In the Fujikawa method, one chooses the plain-wave eigenstates of the free theory as the complete set states 
when one takes the trace.  Therefore, let us consider $H_{DW}^2$ in the free theory
\begin{eqnarray}
  H_{DW}^2 = -\partial_4^2 - \sum_{i=1}^3 \partial_i^2 + M^2 - 2M\gamma_4\delta(x_4),
\end{eqnarray}
assuming the form of its solution as $\phi_{\mathbf{p},\uparrow\downarrow}^{\rm 3D}(\mathbf{x})\otimes\varphi_{\pm}(x_4)$
where $\varphi_\pm (x_4)$ satisfies
\begin{eqnarray}
  \label{eq:EOM2DW}
  (-\partial_4^2 + \vec{k}^2+M^2\mp 2M\delta(x_4))\varphi_\pm (x_4)=\Lambda^2 \varphi_\pm (x_4),
\end{eqnarray}
and $\phi_{\mathbf{p},\uparrow\downarrow}^{\rm 3D}(\mathbf{x})$ is defined as
$\phi_{\mathbf{p},\uparrow\downarrow}^{\rm 3D}(\mathbf{x}) = v_{\uparrow\downarrow} e^{i\mathbf{p}\cdot\mathbf{x}}$
with $v_\uparrow = \left(\begin{array}{c}1\\
0
\end{array}
\right)
, 
v_\downarrow = \left(
\begin{array}{c}
0\\
1
\end{array}
\right),
$
and $\tau_3\varphi_\pm (x_4)=\pm \varphi_\pm (x_4)$.
Note that the eigenvalue of $\tau_3$ corresponds to that of $\gamma_4$.

The solutions to Eq.~(\ref{eq:EOM2DW}) are obtained as
\begin{eqnarray}
  \varphi^\omega_{\pm,o} (x_4) &=& \frac{u_\pm}{\sqrt{4\pi}}\left(e^{i\omega x_4}-e^{-i\omega x_4}\right),\nonumber\\
  \varphi^\omega_{\pm,e} (x_4) &=& \frac{u_\pm}{\sqrt{4\pi(\omega^2+M^2)}}
  \left\{(i\omega\mp M)e^{i\omega |x_4|}+(i\omega\pm M)e^{-i\omega |x_4|}\right\},\nonumber\\
  \varphi^{\rm edge}_{+,e} (x_4) &=& u_+\sqrt{M}e^{-M|x_4|},
\end{eqnarray}
where $\omega=\sqrt{\Lambda^2-\lambda^2-M^2}$, and the subscripts $e,o$ denote even and odd components
under the time reversal $T:x_4\leftrightarrow -x_4$.



\subsubsection{The first term of Eq.~(\ref{eq:goalDW})}
Inserting the complete set of states in the free theory, and computing the trace for general gauge field 
background just in the usual Fujikawa's method, we obtain the first term in Eq.~(\ref{eq:goalDW}) as
\begin{eqnarray}
\lim_{s\to 0}\frac{1}{\Gamma\left(\frac{1+s}{2}\right)}\int_0^\infty dt\; t^{\frac{s-1}{2}}
\lim_{M\to \infty}{\rm Tr}\left[\gamma_5\epsilon(x_4) e^{-t \frac{H_{DW}^2}{M^2}}\right]
&=&\frac{1}{32\pi^2}\int d^4x\epsilon(x_4)
\epsilon_{\mu\nu\rho\sigma}{\rm tr}_{c}F^{\mu\nu}F^{\rho\sigma}.
\end{eqnarray}

\subsubsection{The third term of Eq.~(\ref{eq:goalDW})}
Noticing that the Chern-Simons term appears as the surface contribution at $x_4=0$,
\begin{eqnarray}
  \frac{1}{32\pi^2}\int_{x_4>0} d^4x
  \epsilon_{\mu\nu\rho\sigma}{\rm tr}_{c}F^{\mu\nu}F^{\rho\sigma} &=& \frac{1}{2\pi}CS|_{x_4=0} +\mbox{integer},\\ 
   \frac{1}{32\pi^2}\int_{x_4<0} d^4x
  \epsilon_{\mu\nu\rho\sigma}{\rm tr}_{c}F^{\mu\nu}F^{\rho\sigma} &=& -\frac{1}{2\pi}CS|_{x_4=0} +\mbox{integer}, 
\end{eqnarray}
we can compute the third term in Eq.~(\ref{eq:goalDW}) as
\begin{eqnarray}
-\int_0^1 du \frac{d}{du}(\frac{1}{\pi}CS^u|_{x_4=0}) = -\frac{1}{\pi}CS|_{x_4=0}, 
\end{eqnarray}
where $CS^u$ means the Chern-Simons term with the gauge field $uA_\mu$.

Here the Chern-Simons term is a local part of the
$\eta$-invariant and does not keep the gauge invariance.
In order to keep a manifest gauge invariance,
we can simply add an integer term
\begin{eqnarray}
  \label{eq:etaCS}
\eta(iD^{\rm 3D}) = \frac{1}{\pi}CS|_{x_4=0} - 2\left[\frac{CS|_{x_4=0}}{2\pi}\right],
\end{eqnarray}
where $[f]$
denotes the Gauss symbol or the greatest integer less than or equal to $f$.
In our paper \cite{Fukaya:2017tsq}, we exactly computed
the eta-invariant in one-dimensional $U(1)$ theory
and confirmed that the above simple structure is true.
In the following, we assume that the prescription
with the Gaussian symbol is valid in general gauge theories with three-dimensions.

\subsubsection{The second term of Eq.~(\ref{eq:goalDW})}
For the second term in Eq.~(\ref{eq:goalDW}) requires some technical calculations. 
However, one finds that it only gives contributions which vanishes in the limit.
$s\rightarrow 0$. The second term is shown to be zero. 
For details, see ~\cite{Fukaya:2017tsq}.

\subsubsection{Final result of $\eta(H_{DW})$}
Summing up all the contributions, we obtain
\begin{eqnarray}
\label{eq:etaDW}
  \eta(H_{DW}) = \frac{1}{32\pi^2}\int d^4x\;\epsilon(x_4)
  \epsilon_{\mu\nu\rho\sigma}{\rm tr}_{c}F^{\mu\nu}F^{\rho\sigma}
  -\eta(iD^{\rm 3D}),
\end{eqnarray}
where we simply replaced the Chern-Simons term by the eta-invariant,
adding $2[CS|_{x_4=0}/2\pi]$.
The first term of Eq.~(\ref{eq:etaDW}) contains contribution only from the bulk modes;
the edge-localized modes contribute to $g(x_4,M)$, which disappears in the large $M$ limit,
whereas the second term of Eq.~(\ref{eq:etaDW}) entirely comes from the edge-localized modes,
as explicitly computed in the previous subsection.
Moreover, we can also show
that the index is stable against
any variational changes in $M$ and gauge field $A_\mu$.
It only allows discrete jumps by an even integer in
the boundary contribution $\eta(iD^{\rm 3D})$. 

Together with the Pauli-Villars contribution $\eta(H_{PV})$, we finally obtain
\begin{eqnarray}
  \frac{\eta(H^{reg}_{DW})}{2} &=& \frac{\eta(H_{DW})}{2}-\frac{\eta(H_{PV})}{2}
  =\frac{1}{32\pi^2}\int_{x_4>0} d^4x\;
  \epsilon_{\mu\nu\rho\sigma}{\rm tr}_{c}F^{\mu\nu}F^{\rho\sigma}
  -\frac{\eta(iD^{\rm 3D})}{2},
\end{eqnarray}
which coincides with the APS index.

\section{Summary}

In this work, we have tried to describe the APS index theorem
in a ``physicist-friendly'' way in a simple set-up with a flat metric,
for the Dirac fermion operator with $U(1)$ or $SU(N)$ gauge field background.
Our method corresponds to a generalization of the Fujikawa method on closed
manifolds to that on manifolds with boundaries.

We have defined a new index by the eta-invariant of
the four-dimensional domain-wall fermion Dirac operator with its Pauli-Villars regulator.
The kink structure in the mass term automatically forces the
fermion fields to satisfy a boundary condition, which is locally given
and respects the $SO(3)$ rotational symmetry on the surface.
As a consequence, the edge-localized modes appear in the complete set of the free Dirac operator.
We have applied the Fujikawa method to this complete set satisfying
the non-trivial boundary condition.
Since the boundary condition is no longer dependent on gauge fields,
we do not need the  adiabatic approximation.
We have obtained an index, which
is stable against the changes of mass and gauge field.
This new index coincides with the APS index.

Generalizations to odd-dimensions as well as to other domain-wall systems where
the kink structure is found in other operators than simple mass terms
are also interesting applications. Our study will also be useful in understanding the Zumino-Stora anomaly 
inflow in terms of doubly gapped domain-wall fermion in six-dimensions~\cite{Fukaya:2016ofi}.
\\

We thank K. Hashimoto for organizing the study group on topological
insulators and useful discussions.
We also thank S.~Aoki, M.~L\"uscher, T.~Misumi, H.~Suzuki and A.~Tanaka for discussions.

\bibliography{lattice2017}

\end{document}